\documentclass[sigconf]{acmart}

\usepackage{enumitem}
\usepackage{cleveref}
\usepackage{xcolor, colortbl}
\usepackage{csquotes}
\usepackage{makecell}
\usepackage{multirow}
\usepackage{booktabs}
\usepackage{balance}
\usepackage{dblfloatfix}

\usepackage{diagbox}

\usepackage[disable]{todonotes} %

\usepackage{subfigure}
\graphicspath{{figs/}}

\usepackage[absolute,showboxes]{textpos}

\usepackage{url}

\definecolor{Gray}{gray}{0.8}

\let\orgautoref\autoref
\renewcommand{\autoref}
{\def\sectionautorefname{\S}%
\def\subsectionautorefname{\S}%
\def\subsubsectionautorefname{\S}%
\orgautoref}

\usepackage{pifont}
\newcommand{\cmark}{\ding{51}}%
\newcommand{\xmark}{\ding{56}}%

\usepackage{xspace}
\newcommand{\etal}{\textit{et al.}~}
\newcommand{\eg}{\textit{e.g.,}~}
\newcommand{\ie}{\textit{i.e.,}~}

\newcommand{\one}{({\em i})\xspace}
\newcommand{\two}{({\em ii})\xspace}
\newcommand{\three}{({\em iii})\xspace}

\makeatletter
\renewcommand{\paragraph}[1]{\vspace*{0.03in}\noindent{\bf #1.}\hspace{0.25ex \@plus1ex \@minus.2ex}}
\newcommand{\paragraphND}[1]{\vspace*{0.03in}\noindent{\bf #1}\hspace{0.25ex \@plus1ex \@minus.2ex}}
\makeatother

\copyrightyear{2021}
\acmYear{2021}
\setcopyright{acmlicensed}\acmConference[CoNEXT '21]{The 17th International Conference on emerging Networking EXperiments and Technologies}{December 7--10, 2021}{Virtual Event, Germany}
\acmBooktitle{The 17th International Conference on emerging Networking EXperiments and Technologies (CoNEXT '21), December 7--10, 2021, Virtual Event, Germany}
\acmPrice{15.00}
\acmDOI{10.1145/3485983.3494872}
\acmISBN{978-1-4503-9098-9/21/12}

\begin{document}

\setlength{\TPHorizModule}{\paperwidth}
\setlength{\TPVertModule}{\paperheight}
\TPMargin{5pt}
\begin{textblock}{0.8}(0.1,0.02)
	\noindent
	\footnotesize
	\centering
	If you cite this paper, please use the CoNEXT reference:
	M. Nawrocki, M. Koch, T. C. Schmidt, and M. Wählisch.
	2021. Transparent Forwarders: An Unnoticed Component of the Open DNS Infrastructure.
	\emph{In Proceedings of CoNEXT ’21.}
	 ACM, New York, NY, USA, 9 pages. https://doi.org/10.1145/3485983.3494872
\end{textblock}

\title{Transparent Forwarders: Explaining Unexpected Sources of DNS Responses} %
\title{More Than You Measure: Transparent DNS Forwarders}
\title{Revisiting Public DNS Resolvers (and Forwarders)}
\title[Transparent DNS Forwarders]{Transparent Forwarders: An Unnoticed\\ Component of the Open DNS Infrastructure}

\author{Marcin Nawrocki}
\email{marcin.nawrocki@fu-berlin.de}
\affiliation{%
  \institution{Freie Universit\"at Berlin}
  \country{Germany}
}

\author{Maynard Koch}
\email{maynard.k@fu-berlin.de}
\affiliation{%
  \institution{Freie Universit\"at Berlin}
  \country{Germany}
}

\author{Thomas C. Schmidt}
\email{t.schmidt@haw-hamburg.de}
\affiliation{%
  \institution{HAW Hamburg}
  \country{Germany}  
}

\author{Matthias W\"ahlisch}
\email{m.waehlisch@fu-berlin.de}
\affiliation{%
  \institution{Freie Universit{\"a}t Berlin}
  \country{Germany}
}

\renewcommand{\shortauthors}{Nawrocki, et al.}


\begin{abstract}
In this paper, we revisit the open DNS (ODNS) infrastructure and, for the first time, systematically measure and analyze transparent forwarders, DNS components that transparently relay between stub resolvers and recursive resolvers.
Our key findings include four takeaways.
First, transparent forwarders contribute 26\% (563k) to the current ODNS infrastructure.
Unfortunately, common periodic scanning campaigns such as Shadowserver do not capture transparent forwarders and thus underestimate the current threat potential of the ODNS.
Second, we find an increased deployment of transparent forwarders in Asia and South America.
In India alone, the ODNS consists of 80\% transparent forwarders.
Third, many  transparent forwarders relay to a few selected public resolvers such as Google and Cloudflare, which confirms a consolidation trend of DNS stakeholders.
Finally, we introduce DNSRoute++, a new traceroute approach to understand the network infrastructure connecting transparent forwarders and resolvers.
\end{abstract}

\maketitle

\section{Introduction}
\label{sec:introduction}

\setlength{\tabcolsep}{2.5pt}
\begin{table}
\caption{Comparison of known open DNS components.}
\label{tbl:comparison-forwarders}
\begin{tabular}{lrrrrrr}
\toprule
& \multicolumn{1}{c}{2014} & \multicolumn{1}{c}{2020} & \multicolumn{4}{c}{2021}
\\
\cmidrule(l{2pt}r{2pt}){2-2} \cmidrule(l{2pt}r{2pt}){3-3} \cmidrule(l{2pt}r{2pt}){4-7}
& \cite{kuhrer2014exit} & \cite{niaki2020cache} & \cite{censys2017search} & \cite{shadowserverDNS} & \cite{ShodanWebsite} & This Work
\\
\midrule

\# Rec. Resolvers &
n/a
&
20K %
&
50K %
&
n/a
&
n/a
&
32K (2\%)
\\

Forwarders & \\
\,\, \# Recursive &
n/a
&
1.4M %
&
1.7M %
&
n/a
&
n/a
&
1.5M (72\%)%
\\

\,\, \# Transparent &
0.6M (2\%) & %
n/a
&
n/a
&
n/a
&
n/a
&
0.6M (26\%) %
\\
\midrule

All ODNSes &
25.6M
&
1.42M
&
1.75M
&
1.8M
&
1.6M
&
2.125M
\\
\bottomrule
\end{tabular}
\end{table}

\setlength{\tabcolsep}{4pt}

The open DNS infrastructure (ODNS)~\cite{schomp2013client} comprises all components that publicly resolve DNS~queries on behalf of DNS~clients located in a remote network.
This ``openness'' makes the ODNS~system a popular target for attackers, who are in search for amplifiers of DNS requests, for periodic DNS~scan campaigns, which try to expose the attack surface, and for researchers, who want to learn more about DNS behavior.

Originally observed in 2013~\cite{mauch2013nanog}, \emph{transparent} DNS~forwarders have not been analyzed in detail since then, but fell off the radar in favor of  \emph{recursive} forwarders and resolvers.
This raises concerns for two reasons.
First, the relative amount of transparent forwarders increased from 2.2\%
in 2014 to 26\% in 2021 (see \autoref{tbl:comparison-forwarders}).
Second, as part of the ODNS, they interact with unsolicited, potentially malicious requests.

In this paper, we systematically analyze transparent forwarders. Our main contributions read as follows:
\begin{enumerate}
\item We characterize transparent forwarders and review DNS~measurement methods. (\autoref{sec:related_work})
\item We experimentally show that popular DNS scanning campaigns do not expose transparent forwarders and thus provide an incomplete view on the ODNS threat landscape. (\autoref{sec:censysetal})
\item We measure the impact of transparent forwarders and find diverse deployments, heavily dependent on the hosting country. For example, configurations of forwarders in South America and Asia greatly contribute to DNS~consolidation.~(\autoref{sec:analysis})
\item We introduce \texttt{DNSRoute++}, a new traceroute approach that leverages the behaviour of transparent forwarders and explores interconnectivity in the ODNS. (\autoref{sec:dnsroute})
\item We discuss transparent forwarders in a broader context. Most of the transparent forwarders are CPE devices, either serving single end customers or larger networks. (\autoref{sec:discussion}) 
\end{enumerate}

\section{Background and Related Work}
\label{sec:related_work}

\begin{figure}[t]
  \begin{center}
  \includegraphics[width=1\columnwidth]{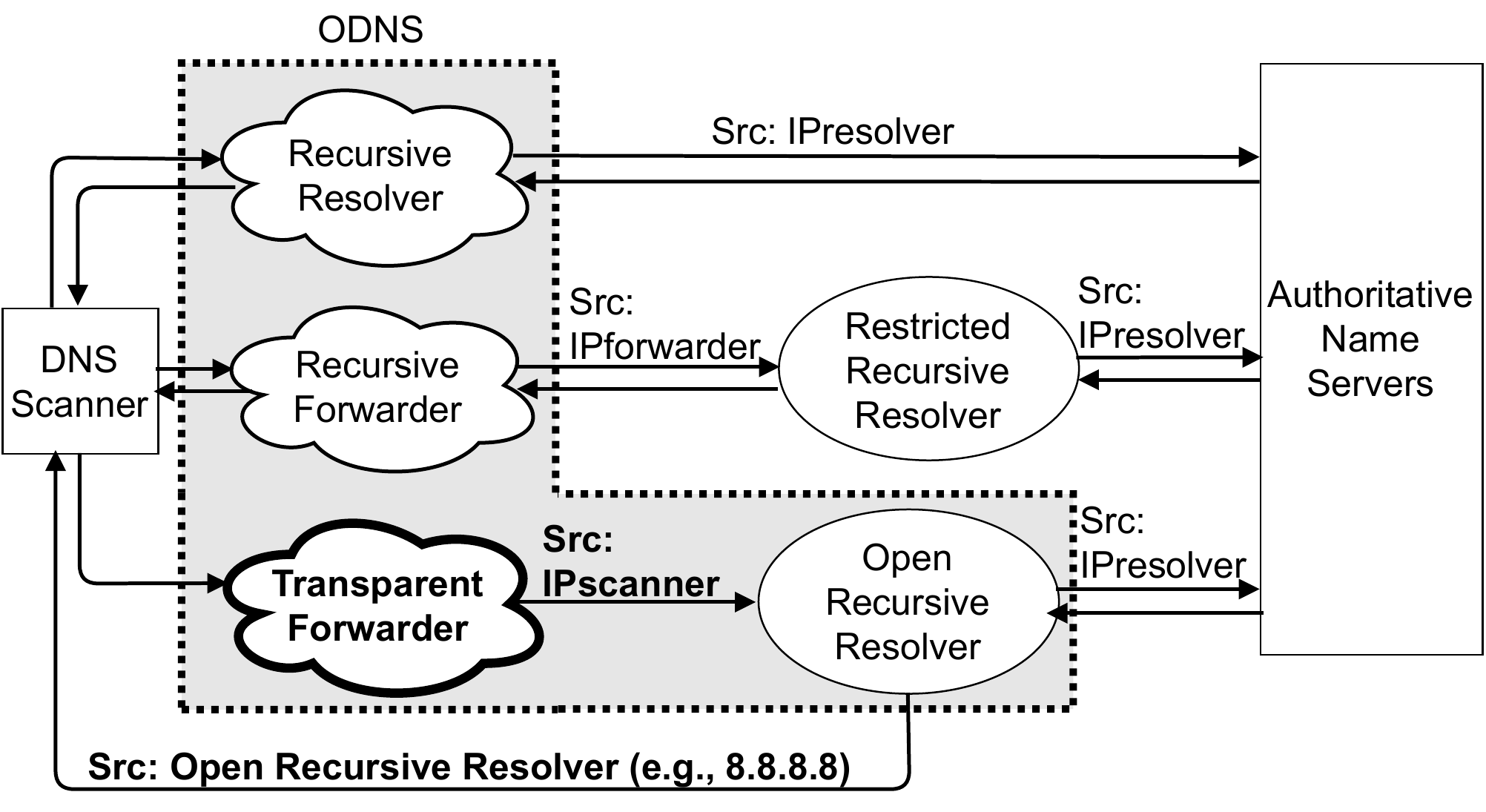}
      \caption{Overview of various ODNS components and their relation to common Internet-wide scan setup.}
  \label{fig:overview_scanning}
  \end{center}
  \vspace{-0.45cm}
\end{figure}

\paragraph{Open DNS (ODNS)}
Various DNS stakeholders \cite{almeida2017stakeholders} such as domain owners and network operators operate autonomously and pursue different goals.
A common view on the DNS is the ODNS~infrastructure~\cite{schomp2013client}, client-side DNS speakers that openly accept requests from any host (not related to oblivious DNS, \ie ODoH).
ODNS components have been previously classified into \emph{recursive resolvers} and \emph{forwarders} \cite{schomp2013client, anagnostopoulos2013dns}.
Recent Internet-wide scans show that the majority (95\%) of ODNSes are forwarders \cite{niaki2020cache} but prior work does not further distinguish between recursive and transparent forwarders and mainly assume only the presence of recursive forwarders \cite{kuhrer2015resolvers}.

\autoref{fig:overview_scanning} shows the expected behavior of all three ODNS components, which are commonly used by stub clients.
Recursive resolvers send queries recursively to authoritative name servers and respond with the final answer to the original client (\eg scanner).
In contrast, forwarders do not use DNS primitives to resolve names but forward queries to a recursive resolver~\cite{RFC-8499}.
Recursive forwarders can relay to restricted resolvers, however, 
transparent forwarders must forward to open resolvers to act as an ODNS (otherwise, the resolver rejects the scanner IP address).
Upon receiving a final answer, a forwarder may cache it and relays it to the client.
Forwarders are often susceptible to fragmentation~\cite{zlpyz-potfc-20} or side-channel~\cite{man2020poisening} attacks.

\paragraph{Introducing Transparent Forwarders}
In this paper, we argue that there are DNS~forwarders that do not receive DNS~answers because they operate completely transparent.
Such deployment makes the distinction between recursive forwarders and transparent forwarders necessary.

A \emph{recursive forwarder} replaces the original source IP~address of the client by its own IP~address.
A \emph{transparent forwarder} keeps the IP address of the original requester (\eg $IP_{Scanner}$).
The relaying behavior of transparent forwarders has two implications.
First, answers are sent back directly from resolvers to the original requester, \ie they are neither observed nor cached by the forwarder.
Second, transparent forwarders spoof the IP~address of the requester.

Surprisingly, Internet-wide, single packet scans lead to multiple answers from the same host, \eg 314k responses from \texttt{8.8.8.8}.
Our study verifiably links these to transparent forwarders.
Prior work removes these in a sanitizing step \citep{niaki2020cache} or describes them as \emph{unexpected} \cite{lu2021unexpected} but falls short to identify the root cause.
So far, transparent forwarders have been treated as an artifact which can be utilized to measure missing outbound source-address-validations~\cite{kuhrer2014exit, korczynski2020sav}.

\paragraph{Comparison of ODNS Classification Methods}
Two methods are common to distinguish recursive resolvers and forwarders:
\one Destination-specific DNS~queries from a scanner, which encode the destination IP addresses as a subdomain into the query name (\eg \texttt{203-0-113-1.mydomain.com}).
\two Source-specific responses from an authoritative name server, which inserts the IP address of the  immediate client (\eg \texttt{203.0.113.1}) into a dynamic \texttt{A}~resource record of the query name (\eg \texttt{mydomain.com A 203.0.113.1}).
This method can utilize two \texttt{A}~resource records, a client-specific record and a static control record to check for DNS manipulations.

The first method enables an analysis at the name server.
If the IP source address of an immediate client matches the encoded IP address within the query name, then the scanned destination is a recursive resolver, and a forwarder otherwise.
The second method requires an  analysis at the node that originally sent the query (\eg a DNS scanner).
If the IP source address of the response matches the IP address within the \texttt{A} record, then the scanned destination is a recursive resolver, otherwise the scanned node is a DNS~forwarder.
This condition does not hold true anymore for transparent forwarders as recursive resolvers reply directly to the scanning node.

\autoref{tbl:comparison-classification} summarizes the (dis-)advantages of both methods.
The query-based method is particularly useful when the measurement objective needs to prevent caching, because the query name is unique for each target.
This increases the load at the authoritative name server, though.
The response-based method keeps the load at the authoritative name server low since it allows to utilize caches.
Although the first method allows to detect forwarders already at the name server, classifying forwarders into recursive and transparent is only possible at the scanning node because the source IP address of the response has to be evaluated. %
Such a classification requires a correlation of DNS requests and responses to reflect the full DNS~transaction.
Hence, we will deploy the latter method in~\autoref{sec:analysis}.

\begin{table}
\caption{Comparison of forwarder detection methods.}
\label{tbl:comparison-classification}
\begin{tabular}{lrr}
\toprule
& \multicolumn{2}{c}{Custom}
\\
& Queries & Responses \\
& \cite{allman2020dnscontext, schomp2013client, kuhrer2014exit, korczynski2020sav} & \cite{niaki2020cache, censys2017search}, this work\\    
\midrule
    Utilization of caches & None & High \\
    Load auth. name server & High & Low \\
    Forwarder detection & At server & At client \\
    Forwarder classification & At client & At client \\
\bottomrule
\end{tabular}
\end{table}

\section{Popular Scanning Campaigns and Transparent Forwarders}
\label{sec:censysetal}
Censys~\cite{censys2017search}, Shadowserver~\cite{shadowserverDNS}, and Shodan are popular scanning campaigns to reveal ODNSes.
To verify our assumption that these campaigns underestimate the current number of ODNSes because they only record responses without correlating with the original target IP~addresses of requests, we conduct a controlled~experiment.

\subsection{Controlled Experiment}
\label{sec:controlled-experiment}
\begin{figure}[t]
  \begin{center}
  \includegraphics[width=1\columnwidth]{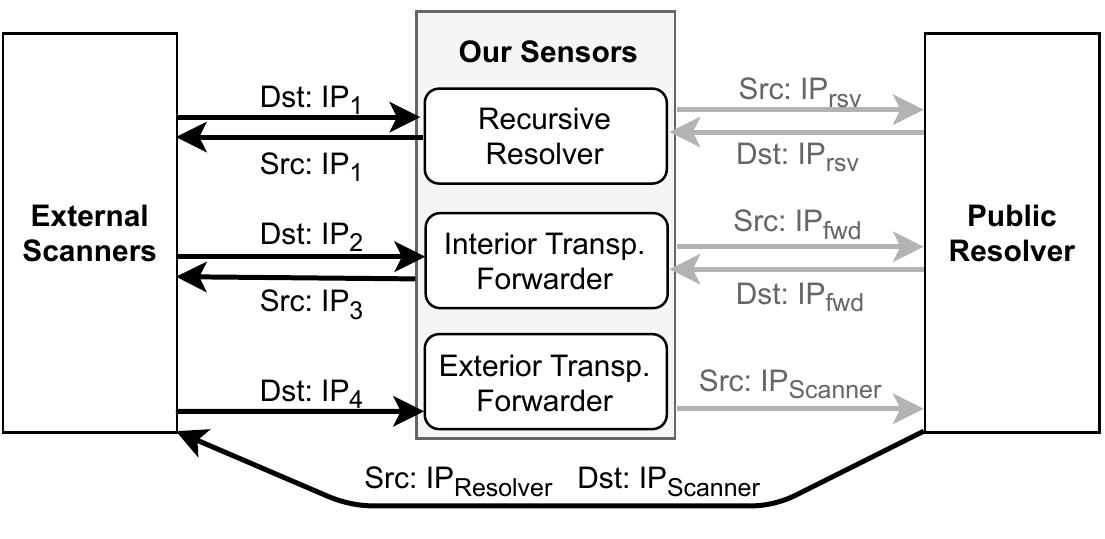}
      \caption{DNS sensors. Black arrows indicate DNS messages visible to external scanning campaigns.}
  \label{fig:overview_honeypots}
  \end{center}
  \vspace{-0.3cm}
\end{figure}

We develop and deploy three ODNS honeypot sensors, see~\autoref{fig:overview_honeypots}.

\paragraph{Sensor 1: Recursive Resolver}
The first sensor behaves exactly like a public recursive resolver.
The sensor answers using the same IP address at which it also has received a DNS request, $IP_1$.
This configuration is a baseline measurement.
We expect every viable Internet-wide DNS scanning campaign to find this sensor.

\paragraph{Sensor 2: Interior Transparent Forwarder}
We utilize two IP addresses, $IP_2$ to receive DNS requests from a scanner and $IP_3$ to send responses. %
Both IP addresses are part of the same \texttt{/24} prefix.
This configuration allows for the following inferences:
\one Scanners that report $IP_2$ ignore the different IP address $IP_3$ in the response.
They are RFC-compliant~\cite{RFC-1035}, and implement DNS transactional scans.
\two Scanners that report $IP_3$ only evaluate the responses independently of the sent requests, which is a strong indicator for stateless, response-based analysis.
This sensor mimics the key behavior of a transparent forwarder, but, as both addresses belong to the same IP~prefix, the setup is easy to deploy.
It does not require special network configuration such as disabled source address validation.
Moreover, we can ensure that a reply is sent to the scanner.

\paragraph{Sensor 3: Exterior Transparent Forwarder}
The third sensor implements a transparent forwarder which relays spoofed packets to an external, public resolver.
This sensor is reachable at $IP_4$ and forwards a request using the source IP~address of the scanner.
To allow for spoofing, this sensor should be connected to a network that does not deploy source address validation~\cite{RFC-2827} and peers directly with the network of the public resolver.
In contrast to the previous setups, we do not receive the answer from the public resolver since the answer is sent directly to the scanner.
Similarly to sensor~(2), we can infer the following:
\one Scanners that report $IP_4$ ignore the different IP address in the response, indicating transactional scans.
\two  Scanners that report the public resolver will miss our forwarding sensor.
This is because multiple responses from the same source will be aggregated into a single DNS speaker. %

\paragraph{Deployment Details}
Our sensors resolve incoming requests using Google's public resolver. %
We verify that source address validation is not deployed in our network.
Moreover, our network peers directly with Google at an Internet eXchange point (IXP), so we are not exposed to filters from upstream providers, as required for sensor~3.
We confirm the correct operation of all sensors by sending DNS queries and analyzing replies at the scanner.
To prevent amplification attacks~\cite{rowrs-adads-15}, we configure a strict rate limiting such that each sensor is allowed to answer one request every 5 minutes per source \texttt{/24} prefix.
We use a rate limiting based on the client prefix since it also prevents DoS carpet bombs \cite{heinrich2021carpet}.
We deploy our sensors for multiple weeks and then inspect the scan project results.

\subsection{Results}
All three sensors received scans from Censys, Shadowserver, and Shodan, but
those scanners did not identify all of our sensors as an ODNS~component.
We use Censys' and Shodans public search API to check which IP~addresses of our sensors have been discovered.
As owner of the IP~prefix that we used for our sensors, we have been informed by Shadowserver 
about our sensors.

All measurement campaigns discovered our public resolver (Sensor~1).
None of them found one of our DNS~forwarders, see \autoref{tbl:sensors-detected}.
Shadowserver reported the replying IP~address $IP_3$ of Sensor 2, which, in real deployment, would represent the address of a recursive resolver.
However, Censys and Shodan did not report $IP_3$, which indicates that the responses did not pass a sanitizing step, respectively. %
We conclude that transparent forwarders are currently missed by these scanning campaigns.
Given that the measurement results of these campaigns are used by third parties, the impact of ignoring transparent fowarders is large.
National CERTS, for example, rely on data from Shadowserver to identify local ODNS~systems.

\setlength{\tabcolsep}{2pt}
\begin{table}
\caption{\mbox{Detection of our DNS~sensors by popular scans.}}
\label{tbl:sensors-detected}
\begin{tabular}{lcccc}
\toprule
 & \multicolumn{4}{c}{Detected} \\
 \cmidrule{2-5}
& Sensor~1 & \multicolumn{2}{c}{Sensor~2} & Sensor~3
\\
\cmidrule{2-2}
\cmidrule(lr){3-4}
\cmidrule(l){5-5}
Scanner & $IP_1$ & \ \ $IP_2$ & $IP_3$ & $IP_4$ \\
\midrule
Shadowserver 		& \cmark & \xmark & \cmark & \xmark \\
Censys 	& \cmark & \xmark & \xmark & \xmark \\
Shodan 	& \cmark & \xmark & \xmark & \xmark \\
\bottomrule
\end{tabular}
\vspace{-0.2cm}
\end{table}
\setlength{\tabcolsep}{4pt}

\section{Measuring and Analysing Transparent DNS Forwarders}
\label{sec:analysis}

\subsection{Measurement Method and Setup}
\label{sec:method-setup}
\paragraph{Method}
To identify transparent forwarders, we need to correlate requests and responses at the scanning node.
Our method aims for easy deployment, low measurement overhead, and robustness against manipulations.
It requires two steps.
First, mapping replies to requests of our scans.
Second, classifying ODNS components.

To implement the first step, our scanner records the complete DNS transaction, \ie source and destination IP~addresses, client port, and the ID used in the DNS header~\cite{RFC-1035}.
Assigning replies to requests based only on IP addresses would introduce ambiguity since replies triggered via transparent forwarders will include the source IP~address of the resolver.
Furthermore, to enable Internet-wide parallel scans, we ensure unique tuples of transport port and ID similar to other asynchronous scanners~\cite{durumeric2013zmap}.
Then, even if we receive replies from the same resolver used by different transparent forwarders, we can clearly map responses to requests (for a detailed example, compare appendix \autoref{fig:overlap_transparent}).

\setcounter{figure}{3}
\begin{figure*}[!b]
  \begin{center}
  \includegraphics[width=2.0\columnwidth]{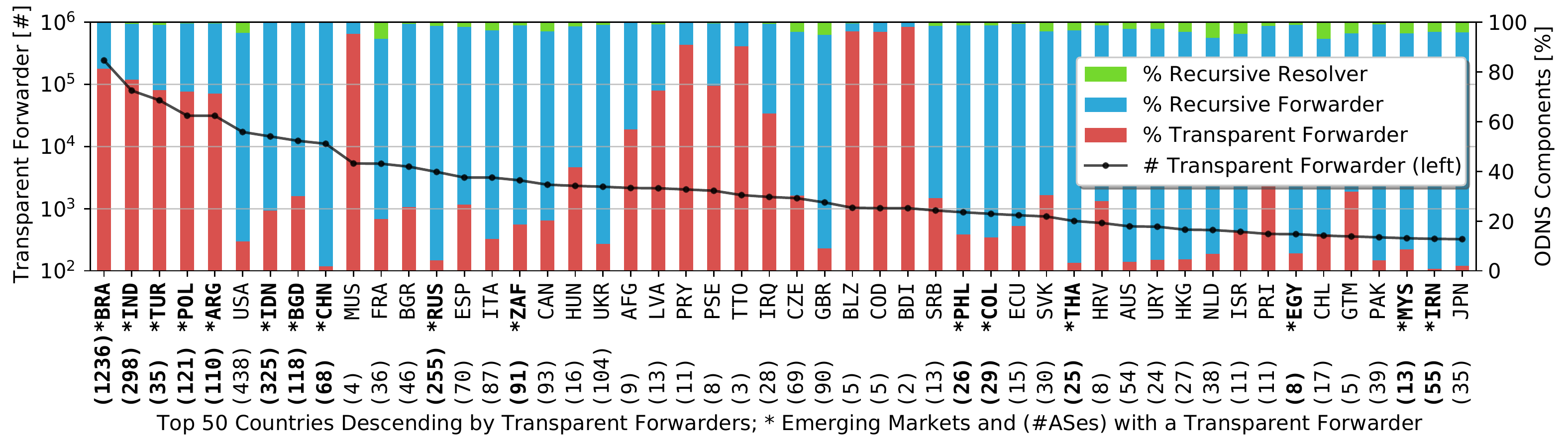}
      \caption{\mbox{Top-50 countries with transparent forwarders.
    Countries with emerging markets exhibit more transparent forwarders.}}
  \label{fig:topxcc_odns_shares}
  \end{center}
\end{figure*}

Our scanner requests a static name that belongs to a DNS zone which we control.
The corresponding authoritative name server replies with two \texttt{A} records similar to other approaches using client-specific responses (details see \autoref{sec:related_work}).
Performing full DNS transactions and using a control resource record also helps us to identify distortions introduced by middleboxes~\cite{itd-liuis-21}.
After receiving replies, we correlate the client port number and DNS transaction ID of responses and previously recorded request data.
We use a conservative DNS~timeout of 20 seconds. %
Note that this and the subsequent analysis of forwarders is part of post-processing the data.
It does not affect the speed of scanning.

We then classify ODNS components.
Utilizing the destination address of the request ($IP_{target}$), the response source address ($IP_{response}$) and dynamic \texttt{A} resource record ($A_{resolver}$), we apply:

\begin{description}
  \item[Transparent Forwarder] if 
  \newline 
  $ IP_{target} \neq IP_{response} $ %

  \item[Recursive Forwarder] if
  \newline 
  $ IP_{target} = IP_{response}\, \land\, IP_{response} \neq A_{resolver} $

  \item[Recursive Resolver] if
  \newline 
  $ IP_{target} = IP_{response} \, \land \, IP_{response} = A_{resolver} $
\end{description}

\paragraph{Setup}
We deploy our scanner in a network, which allows for high packet rates without triggering a DoS attack mitigation such as packet drops or rate limiting.
We probe any public IPv4~address and use moderate scanning rates, \ie we need 18 hours for a full pass.
Our authoritative name server is implemented based on a common high-performance DNS library \cite{gieben2010godns}, which supports up to 20k pps.

\subsection{Results}

The subsequent results are based on an Internet-wide scan from April~20,~2021.
Ongoing, more recent scans find the same results.

\paragraph{Detailed Comparison with Shadowserver}
We find $\approx$536k transparent forwarders, identified by distinct IP addresses.
Compared to Shadowserver~\cite{shadowserverDNS}, which does not detect transparent forwarders, this reveals $\approx18\%$ more ODNS components (compare \autoref{tbl:comparison-forwarders}).

It is worth noting that we identified, in sum, fewer recursive resolvers and recursive forwarders compared to Shadowserver, because we require responses to include both \texttt{A}-records, with the static control record being unaltered.
Shadowserver requires only one correct \texttt{A} record.
Omitting this step in our method leads to similar numbers than Shadowserver.
To be robust against manipulation, we keep our more strict requirement and still detect more ODNS components in total due to consideration of transparent forwarders.

\setcounter{figure}{2}
\begin{figure}[t]
  \begin{center}
  \includegraphics[width=1\columnwidth]{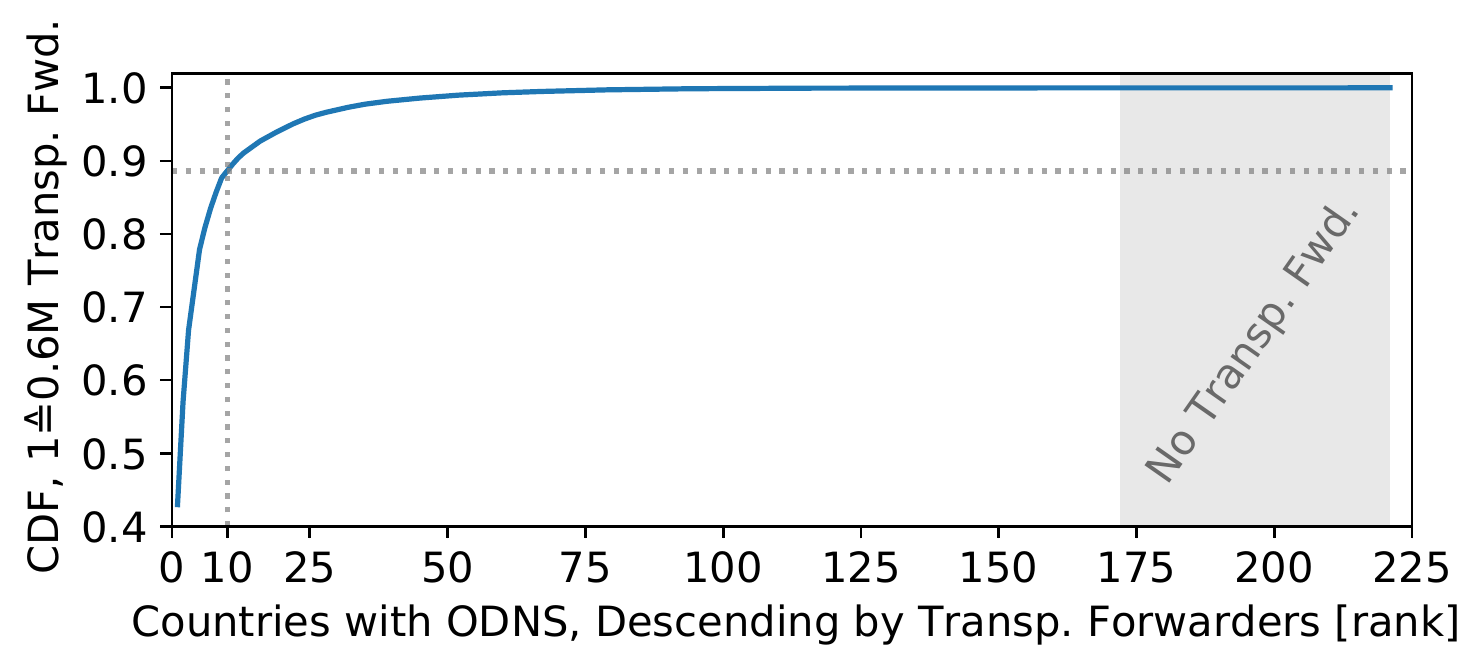}
      \caption{CDF of transparent forwarders per country. Top-10 countries exhibit $\sim$90\% of all transparent forwarders.}
  \label{fig:transp_fwd_count}
  \end{center}
  \vspace{-0.50cm}
\end{figure}
\setcounter{figure}{4}

\begin{figure*}[t]
  \begin{center}
  \includegraphics[width=2.0\columnwidth]{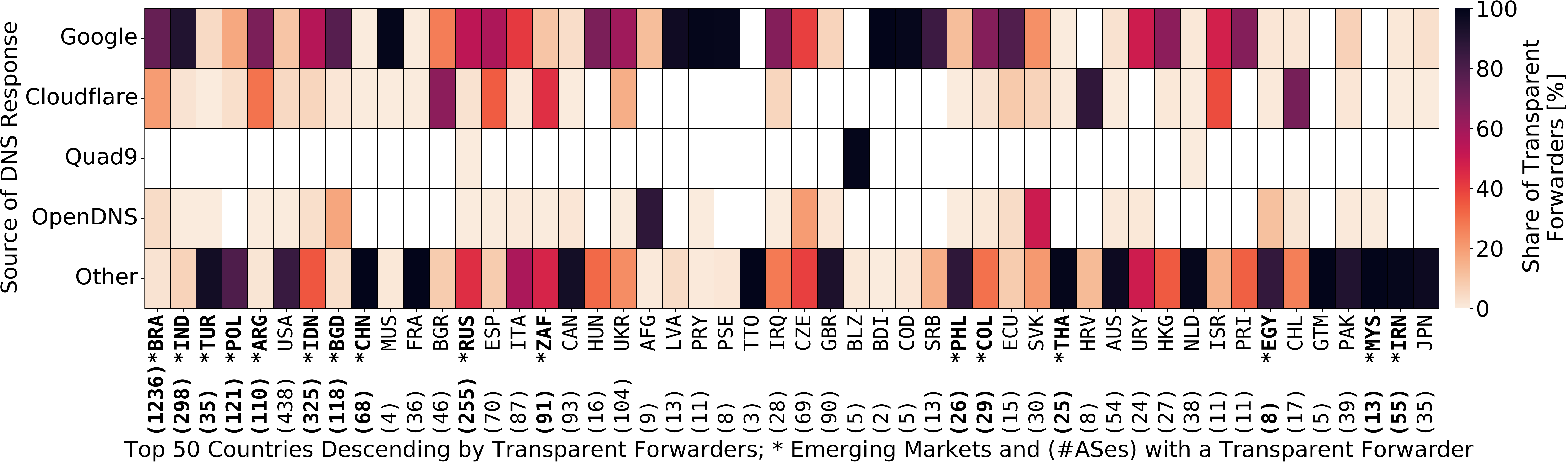}
      \caption{Popularity of public resolver projects. Google \& Cloudflare are commonly used by transparent forwarders.}
  \label{fig:heatmap_dominance_pub_rr}
  \end{center}
\end{figure*}

\paragraph{Geo-Location of Transparent Forwarders}
We now try to understand whether the deployment of transparent forwarders is more popular in specific countries. %
To this end, we successfully map 99.9\% IP~addresses to ASes based on Routeviews dumps. 
Then, we map ASes to country codes with \textit{whois} data und MaxMind. 
\autoref{fig:transp_fwd_count} depicts the cumulative number of forwarders per country.
Roughly 25\% of countries with at least one ODNS component do not exhibit any transparent forwarder (highlighted in gray).
We find, though, that ten~countries host 90\% of all transparent forwarders .

Countries that only expose transparent forwarders to the ODNS may be missed completely by scanning campaigns.
Considering our complete data set, we do not find those cases.
We find 5 countries hosting over 90\% transparent forwarders, 4 of them are among the top-50 countries (see \autoref{fig:topxcc_odns_shares}).
8 out of 9 countries with over 10k transparent forwarders are classified as an emerging market \cite{casanova2020emerging}, such as Brazil and India.
With respect to all ODNSes in these two countries, transparent forwarders account for more than 80\%.

\paragraph{Common Public Resolvers used by Transparent Forwarders}
DNS consolidation directly correlates with how difficult it is to detect transparent forwarders.
This is because the higher the consolidation, the more forwarders are \emph{hidden} behind individual resolver IP addresses. %
Hence, we analyze the used resolvers and assess the relative popularity of four large public resolver projects (Google, Cloudflare, Quad9, and OpenDNS) per country. %
\autoref{fig:heatmap_dominance_pub_rr} unveils that Google and Cloudflare are most common.
Almost all transparent forwarders in India relay requests to Google, for example.
This aligns with recent complementary studies, which show that 19\% of Google DNS users are located in India~\cite{ripe82-huston-dns}.
Following these results we can conclude that current scanning campaigns, which only consider DNS~replies, underestimate the amount of ODNS components per country since they observe responses only from \texttt{8.8.8.8} or other public DNS projects.
Comparing Shadowserver and our data, the ODNS~rank of the top-20 countries varies up to 12 positions (details see \autoref{apx:ranking}).

\setlength{\tabcolsep}{5pt}
\begin{table}[b]
    \caption{Top-10 countries with highest ``other'' share (absolute) in Fig.~\ref{fig:heatmap_dominance_pub_rr}. We show  \one the ASNs from which our scanner received most of the ``other'' responses, \two the number of transparent forwarders, \three the share of responses in ``other'' for which the ASN of $A_{resolver}$ belongs to one of the four common resolver~projects.}
    \label{tbl:share_of_other_portion}
    \begin{tabular}{lrrr}
        \toprule
        Country & Top ASN & \thead{\# Transparent \\ Forwarders} & \thead{Indirect \\ Consolidation}\\
        \midrule
        Turkey & 9121 & 52,663 & 0.3\% \\
        Poland & 5617 & 24,879 & 1.4\% \\
        United States & 209 & 14,546 & 18\% \\
        China & 4812 & 11,030 & 0.9\% \\
        France & 5410 & 5,268 & 0.8\% \\
        Indonesia & 4622 & 5,154 & 27\% \\
        India & 3356 & 5,037 & 48\% \\
        Brazil & 262462 & 4,920 & 48\% \\
        Canada & 21724 & 2,303 & 21\% \\
        Italy & 3269 & 1,824 & 35\% \\
        
        \bottomrule
    \end{tabular}
\end{table}

\paragraph{Alternative Resolvers used by Transparent Forwarders}
We find countries in which transparent forwarders do not use one of the four common resolver projects (see ``other'' in \autoref{fig:heatmap_dominance_pub_rr}).
In order to understand the usage of alternative resolvers, we analyze the top-10 countries with most transparent forwarders in the  ``other'' share. %
Our results are summarized in \autoref{tbl:share_of_other_portion}.
We detect two trends.
First, countries such as India and Italy that already use popular resolver projects frequently (direct DNS~consolidation) also deploy complex forwarding chains.
In those cases, at our scanner, we receive DNS responses from IP addresses belonging to the AS of the transparent forwarder.
Analysing the IP address in the $A_{resolver}$ record reveals, however, that our authoritative name server received the request from an IP address outside this AS.
This unveils a dependency chain in which transparent forwarders relay to local recursive forwarders, which then forward to a popular resolver project (indirect consolidation).
Second, we identify countries (Poland, France, China, and Turkey) that tend to not use public resolvers at all.
Here, we find larger forwarder pools but those forwarders use only 1 to 10 local resolvers.
For example, a single DNS resolver (195.175.39.69, Turkish Telecom) is serving almost all transparent forwarders from Turkey, which again masks their existence (for stateless scans).

\vspace{-0.25cm}
\section{DNSRoute++}
\label{sec:dnsroute}

In this section, we introduce \emph{DNSRoute++}, a tool to explore network properties around transparent forwarders, and present two results. %

\paragraph{Measurement Approach}
\emph{DNSRoute++} is a traceroute application that exploits the behavior of transparent forwarders.
In contrast to common \texttt{traceroute}, \emph{DNSRoute++} sends DNS requests and continues incrementing the TTL when the target is reached.
If the target IP~address is a transparent forwarder, we expect to receive \texttt{TTL Exceeded} messages from hosts beyond the forwarder.
In detail, \emph{DNSRoute++} \one reveals all hops between a scanner and the (target) transparent forwarder, then \two all hosts between the transparent forwarder and the recursive resolver used by the forwarder.
This works because the IP~stack of the transparent forwarder replies when the TTL is exceeded (which stops forwarding) and forwards a DNS~request internally to the upper layers otherwise (which reveals hosts beyond a forwarder).
We scan all transparent forwarders.

\begin{figure}[t]
  \begin{center}
  \includegraphics[width=1\columnwidth]{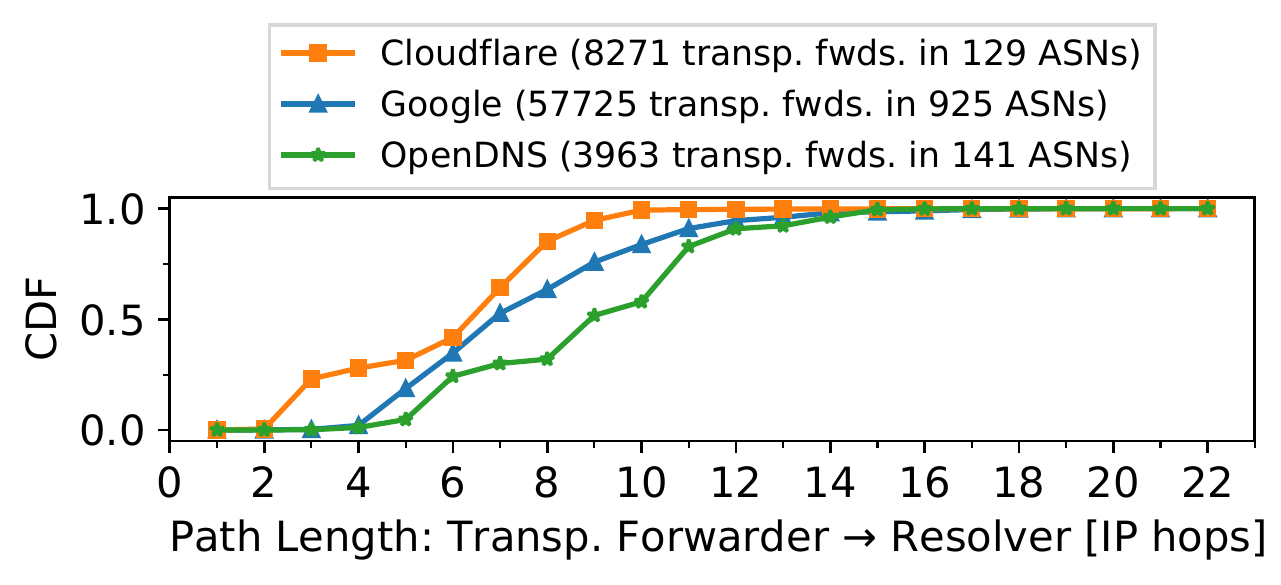}
      \caption{Distribution of path lengths between transparent forwarders and their recursive resolvers, separated by common recursive resolver projects.}
  \label{fig:transp_fwd_hops_resolver}
  \end{center}
  \vspace{-0.30cm}
\end{figure}

\paragraph{Path Lengths to Public Resolvers}
\mbox{We~compare~path~lengths~from} transparent forwarders to their recursive resolvers, see \autoref{fig:transp_fwd_hops_resolver}.
We obtain over 70k~paths to 1.1k ASNs after sanitization.
Our sanitization removes incomplete paths due to host churn or traceroute anomalies.
Short path lengths indicate sound anycast~deployments.

We find that Cloudflare exhibits the shortest paths compared to Google and OpenDNS.
On average, Cloudflare resolvers are reachable in 6.3~hops.
In case of Google and OpenDNS, we observe 7.9 and 9.3~hops, respectively.

Doan \etal \cite{doan2021centralization} performed similar path measurements using 2.5k~RIPE Atlas probes in 729 distinct ASes.
They also observe shorter paths to the Cloudflare resolver but a reverse ranking in case of Google and OpenDNS.
This difference might be due to the location of measurement probes.
RIPE Atlas probes are more likely located in North America and Europe, transparent forwarders are more common in South America and Asia.
It is worth noting that our measurement approach only requires transparent forwarders and no deployment of dedicated probes in external networks.
Hence, our methodology is complementary.

\paragraph{AS Relationship Inference}
Paths~acquired~with~\emph{DNSRoute++}~may help to infer AS relationships.
The \mbox{autonomous system~(AS)~before} the AS of a forwarder indicates an inbound network ($AS_{in}$) and the AS after a forwarder the outbound network~($AS_{out}$).
If $AS_{in}=AS_{out}$, we can assume a provider-customer relationship, since our scanner is outside the customer cone of $AS_{in}$.
After sanitizing AS mappings,  we can utilize 27k paths and observe $AS_{in}=AS_{out}$ for 62\% of the paths.
We detect 41~provider-customer relationships that are currently unclassified by
CAIDAs relationship inference~\cite{caida_asrank}.

\section{Discussion}
\label{sec:discussion}

\paragraphND{What is the purpose of transparent forwarders?}
Transparent forwarders differ from intentional DNS manipulations.
First, they are not part of transparent interception \cite{randall2021homejack, jones2016root, liu2018interception, wei2020spoofing}, which forwards queries to alternate resolvers and spoofs responses.
Also, they differ from DNS redirection, which changes response records for the sake of advertisement \cite{weaver2011redirecting, kreibich2010netalyzr} or censorship \cite{kuhrer2015resolvers, anonymous2014greatfirewall}.
Lastly, they are not part of DNS tunneling, which carries ancillary information~\cite{ishikura2021tunneling} not related to name resolution.

We conjecture that transparent forwarders are misbehaving CPE devices, either serving a single end customer or larger networks.
To support this hypothesis, we perform an \one~AS-based, \two~device-based, and \three~prefix-based classification.
For details about the classification, we refer to \autoref{apx:prefix}.

Considering the top-100 ASes by the number of transparent forwarders, we find 79\% ASes are eyeball ISPs, 7\% of other types, and 14\% remain unclassified.
65~ASNs are 32-bit numbers~\cite{RFC-4893}, \ie belong to more recent AS  deployments. %

MikroTik produces low-cost routers and CPE devices which are often affected by vulnerabilities \cite{ceron2020mikrotik, baines2019routeros} and have been previously identified as DNS forwarders \cite{kuhrer2014exit}.
MikroTik's price policy seems to attract countries with emerging markets \cite{casanova2020emerging}.
Overall, we attribute about 18k hosts (23\%) to MikroTik.

Finally, we find that 26\% of transparent forwarders are located in a \texttt{/24}~IP~prefix that hosts less than 25~transparent forwarders.
Such sparse population indicates that those forwarders belong to CPE devices (\eg home gateways) of individual end customers.
On the other hand, we also find that 36\% of the transparent forwarders cover a \texttt{/24} network completely.
50\% of the MikroTik routers we identified can be assigned to such a scenario.

All these observations strongly indicate that most of the transparent forwarders are misconfigured CPE devices.
Whether these devices serve as a middlebox for a single customer or as router for a larger network does not affect our results regarding consolidation and attack potential.

\paragraphND{Should scanning campaigns deploy transactional scans?}
Yes.
Based on our measurements, current implementations of stateless DNS scans miss transparent forwarders, which account for 26\% of all ODNS components.
Interestingly, some countries host a disproportional amount of transparent forwarders which makes them far more exposed to misuse than previously assumed.
For those 15~countries, we find that they host at least 50\% of transparent forwarders and twice as much ODNSes as comparable studies detect.

Our transactional scans show that revealing transparent forwarders does not conflict with fast, stateless scans.
Transactional scans require little-to-none changes to existing scanning infrastructures.
Required changes include
\one the recording of outgoing scan traffic and
\two a lightweight post-analysis, which matches queries and responses based on the client port and DNS transaction ID.
These changes do not impair the scanning rate itself.
We focus on DNS over UDP~\cite{RFC-1035} as we do not expect transparent forwarding for DoT~\cite{RFC-7858} and DoH~\cite{RFC-8484} since their connection-based requests conflict with IP~spoofing.
Also, for benevolent scanning campaigns, we recommend utilizing custom responses and \emph{not} custom queries for a forwarder classification to limit adverse effects.
Encoding the IP addresses of targets leads to cache pollution due to negative caching~\cite{RFC-8499} and cache evictions of popular, legitimately used names, which resembles random-subdomain \cite{feibish2017subdomains} and water-torture~\cite{luo2018water} attacks.
We find resolvers serving $>$40k forwarders, which would introduce $>$40k cache entries to a single~resolver.

\paragraphND{What is the misuse potential?}
Transparent forwarders can be misused as invisible diffusers for reflective amplification attacks as they relay the source IP address of the DNS request as-is.
Hence, spoofed packets (allegedly from the victim) are forwarded with the source~address spoofed by the attacker.
Booters offering DDoS services utilize centralized attack infrastructures to reduce costs and maintenance \cite{santanna2015booters}.
Misusing transparent forwarders
\one~allows to reach multiple PoPs of anycast DNS providers despite their centralized infrastructure (\eg Google allows \texttt{ANY} requests) and
\two~impedes attribution by further obfuscating the origin of spoofed~traffic.

Overall, transparent forwarders likely belong to domestic setups but interact with unsolicited, external requests, which might lead to impaired performance, security risks and liability implications.

\section{Conclusion}
\label{sec:conclusion}

We showed that the open DNS infrastructure comprises transparent forwarders---in addition to its recursive components.
These forwarders intensify the perceived threat potential of the ODNS.
We argue to include them in on-going and future measurements as they account for a relevant impact and share.
Our results bolster current concerns regarding consolidation of the DNS, at least for countries that massively host transparent forwarders.

\begin{acks}
We would like to thank our shepherd Marinho Barcellos and the anonymous reviewers for their helpful feedback.
We are grateful to Markus de Br\"un, Anders K\"olligan, and operators of the LACNIC region for fruitful discussions.
We thank Moritz M\"uller and Martino Trevisan for their feedback on the artifacts.
This work was partly supported by the \grantsponsor{BMBF}{German Federal Ministry of Education and Research (BMBF)}{https://www.bmbf.de/} within the project \grantnum{BMBF}{PRIMEnet}.
\end{acks}

\label{lastpage}

\balance
\bibliographystyle{ACM-Reference-Format}
\bibliography{bibliography, rfcs, own}

\appendix
\section{Artifacts}

This section gives a brief overview of the artifacts of this paper.
We contribute tools to conduct follow-up  measurements as well as raw data and analysis scripts to reproduce the results and figures presented in this paper.

\subsection{Hosting}

All artifacts are available in the following repository:
\begin{itemize}
  \item[]\url{https://github.com/ilabrg/artifacts-conext21-dns-fwd}
\end{itemize}
This public repository provides up-to-date instructions for installing, configuring, and running our artifacts.
We also archive the camera-ready version of our software on Zenodo:
\begin{itemize}
  \item[]\url{https://doi.org/10.5281/zenodo.5636314}
\end{itemize}

\subsection{Description}
This bundle of artifacts includes three tools, each located in a separate directory.
Additionally, we provide wrapper scripts for Internet-wide scans.
Finally, we provide notebooks to replicate the results of this paper.

\begin{description}
\item[\texttt{dnsRoute++/}]
Traceroute implementation that maps paths behind transparent forwarders (see \autoref{sec:discussion}).
We also add an IP hitlist that includes transparent forwarders at the time of our measurements.
Please note that IP churn might have change the current state.

\item[\texttt{dns-honeypot-sensors/}]
Honeypot sensors emulating various Open DNS speakers (ODNS), including transparent forwarders.
These scripts emulate three different types of open DNS speakers (see \autoref{sec:controlled-experiment}).

\item[\texttt{recursive-mirror-auth-server/}]
A DNS nameserver implementation that replies with an \texttt{A} record referring to the IP address of the client that sent a DNS query.
This reveals a recursive resolver (see \autoref{sec:method-setup}).

\item[\texttt{dns-scan-server/}]
DNS scanning with \emph{zmap} and \emph{dumpcap} to capture the complete traffic during the scan.
This artifact produces raw PCAP files required by the analysis scripts (need to be processed first).

\item[\texttt{dns-measurement-analysis/}]
Implementation of our postprocessing and sanitizing method which creates a dataframe that is used for further analyses.
This artifact replicates the results of this paper.

\end{description}

\subsection{Dependencies}

\paragraph{Software}
For a full list of dependencies, please, see the \texttt{readme.md} file included in each directory.
The minimal requirement is a Linux system, GoLang, Python, and Bash.

\paragraph{Network Infrastructure}
Our software requires specific network setups, such as network access without NAT and DNS names under your control.
Again, please compare the \emph{readme} files.

\subsection{Usage and Testing}

Each directory includes a \texttt{run.sh} and a \texttt{test.sh}.
\begin{description}
  \item[\texttt{run.sh}] wraps multiple sub-scripts of each tool and allows for a quick start.
  \item[\texttt{test.sh}] executes tests and provides expected output, which then can be evaluated for correctness.
For more complex setups, the tests are run against our servers.
\end{description}
The artifacts can be tested independently and in any order.

\subsection{Future Measurements}
We plan to continue our experiments.
Future measurement results will be available on \url{https://odns.secnow.net}.

\section{Ethical Concerns}
We presented a method to discover a new type of public DNS forwarders, which may be misused by attackers.
We are in contact with federal security offices to include transparent forwarders in their on-going  measurements that inform network operators about vulnerable devices.

\section{Ranking Countries by ODNS Components}
\label{apx:ranking}

In this work, we showed that transparent forwarders amount to more than 25\% of all ODNS components.
Common ODNS~scan campaigns such as Shadowserver rank countries based on the number of ODNS components but miss transparent forwarders (see \autoref{sec:censysetal}).
\autoref{tbl:comparison-dns-servers-this-work-and-shadowserver} shows the change of ranks for the top-20 countries when considering the complete ODNS infrastructure by including transparent forwarders.

\definecolor{upvoteGreen}{rgb}{0.13,0.78,0}
\definecolor{downvoteRed}{rgb}{0.78,0.13,0}

\setlength{\tabcolsep}{2.5pt}
\begin{table}[t]
\caption{Top-20 countries ranked by number of ODNS components, comparing this work and Shadowserver.}
\label{tbl:comparison-dns-servers-this-work-and-shadowserver}
\begin{tabular}{lrrrrrr}
\toprule
 & \multicolumn{2}{c}{This Work} & \multicolumn{2}{c}{Shadowserver} & \multicolumn{2}{c}{Difference$\Delta$}  \\
\cmidrule(r){2-3} \cmidrule(r){4-5} \cmidrule{6-7}
Country & Rank & \#ODNS & Rank & \#ODNS & Rank & \#ODNS \\
\bottomrule
China & 1 & 632428 & 1 & 717706 & 0 - & \textcolor{downvoteRed}{$85278 \downarrow$}\\
Brazil & 2 & 297828 & 6 & 49616 & \textcolor{upvoteGreen}{$4 \uparrow$} & \textcolor{upvoteGreen}{$ 248212 \uparrow$}\\
United States & 3 & 144568 & 2 & 137619 & \textcolor{downvoteRed}{$1 \downarrow$} & \textcolor{upvoteGreen}{$6949 \uparrow$}\\
India & 4 & 102910 & 8 & 33510 & \textcolor{upvoteGreen}{$4 \uparrow$} & \textcolor{upvoteGreen}{$69400 \uparrow$}\\
Russia & 5 & 93498 & 3 & 102368 & \textcolor{downvoteRed}{$2 \downarrow$} & \textcolor{downvoteRed}{$8870 \downarrow$}\\
Turkey & 6 & 76168 & 18 & 19298 & \textcolor{upvoteGreen}{$12 \uparrow$} & \textcolor{upvoteGreen}{$56870 \uparrow$}\\
Indonesia & 7 & 59972 & 5 &  56319 & \textcolor{downvoteRed}{$2 \downarrow$} & \textcolor{upvoteGreen}{$3653 \uparrow$}\\
South Korea & 8 & 49143 & 4 & 73790 & \textcolor{downvoteRed}{$4 \downarrow$} & \textcolor{downvoteRed}{$24647 \downarrow$}\\
Argentina & 9 & 43648 & 20 & 16974 & \textcolor{upvoteGreen}{$11 \uparrow$} & \textcolor{upvoteGreen}{$26674 \uparrow$}\\
Poland & 10 & 43431 & 10 & 29175 & 0 - & \textcolor{upvoteGreen}{$14256 \uparrow$}\\
Bangladesh & 11 & 40917 & 16 & 22940 & \textcolor{upvoteGreen}{$5 \uparrow$} & \textcolor{upvoteGreen}{$17977 \uparrow$}\\
Taiwan & 12 & 37550 & 7 & 38525 & \textcolor{downvoteRed}{$5 \downarrow$} & \textcolor{downvoteRed}{$975 \downarrow$}\\
Iran & 13 & 36659 & 9 &  33444 & \textcolor{downvoteRed}{$4 \downarrow$} & \textcolor{upvoteGreen}{$3215 \uparrow$}\\
France & 14 & 25320 & 12 & 25763 & \textcolor{downvoteRed}{$2 \downarrow$} & \textcolor{downvoteRed}{$443 \downarrow$}\\
Italy & 15 & 24766 & 14 &  24483 & \textcolor{downvoteRed}{$1 \downarrow$} & \textcolor{upvoteGreen}{$283 \uparrow$}\\
Vietnam & 16 & 21407 & 15 & 24266 & \textcolor{downvoteRed}{$1 \downarrow$} & \textcolor{downvoteRed}{$2859 \downarrow$}\\
Ukraine & 17 & 20780 & 13 & 25307 & \textcolor{downvoteRed}{$4 \downarrow$} & \textcolor{downvoteRed}{$4527 \downarrow$}\\
Thailand & 18 & 19694 & 17 & 20474 & \textcolor{downvoteRed}{$1 \downarrow$} & \textcolor{downvoteRed}{$780 \downarrow$}\\
Bulgaria & 19 & 18443 & n/a & 16239 & \textcolor{upvoteGreen}{$>\!\!1\uparrow$} & \textcolor{upvoteGreen}{$2204 \uparrow$}\\
Germany & 20 & 16243 & 19 & 17788 & \textcolor{downvoteRed}{$1 \downarrow$} & \textcolor{downvoteRed}{$1545 \downarrow$}\\
\bottomrule
\end{tabular}
\end{table}

\section{Measuring Transparent Forwarders}
\label{apx:method}
\begin{figure}[h]
  \begin{center}
  \includegraphics[width=1\columnwidth]{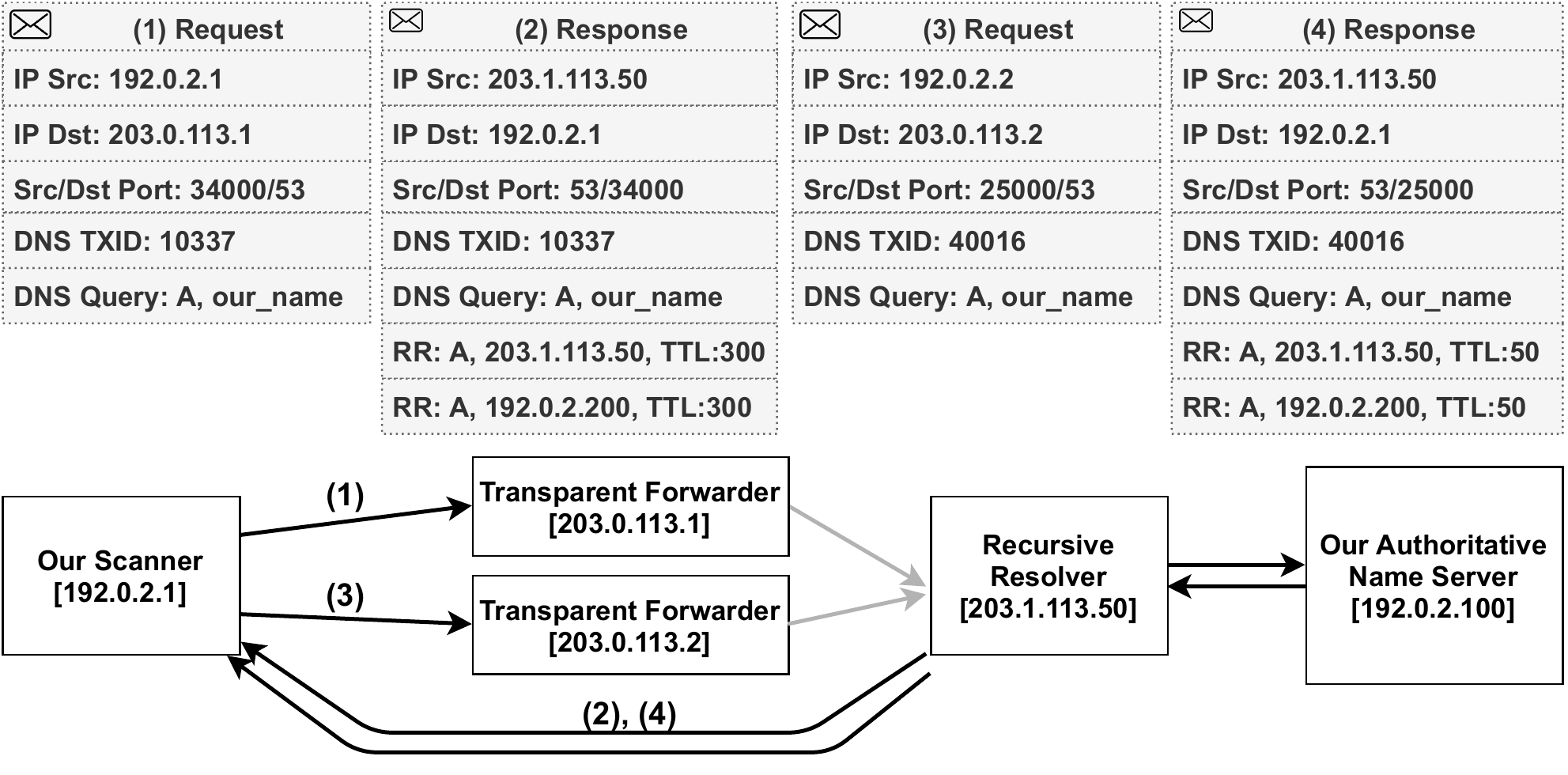}
      \caption{Two transparent forwarders trigger DNS responses from the same recursive resolver, identified by the same source IP address.
      Black arrows indicate DNS messages observed by our infrastructure.}
  \label{fig:overlap_transparent}
  \end{center}
\end{figure}

\autoref{fig:overlap_transparent} illustrates our response-based measurement method, which we explain in detail in \autoref{sec:analysis}.

\section{Details on the Deployment of Transparent Forwarders}
\label{apx:prefix}

\paragraph{AS Classification}
We classify the top-100 ASes by transparent forwarder count.
These ASes cover 50\% of all transparent forwarders.
For each ASN, we map the network type using PeeringDB.
37 ASes are \texttt{Cable/DSL/ISP} networks.
Since the majority of ASes is not classified in PeeringDB, we also perform a manual classification.
We also perform a manual check whether ASes that are classified as \texttt{NSP} provide eyeball Internet services.
Based on our manual inspection, we identify 42 additional ISPs.
In total, out of the top-100 ASes, 79 can be considered \texttt{Cable/DSL/ISP} networks.

\paragraph{Device Fingerprinting}
For device fingerprinting, we use Shodan~\citep{ShodanWebsite} and Censys~\cite{censys2017search}.
To this end, we analyze all open ports and banner grabbing information.
Shodan provides information for 80k of 600k queried hosts. 
Inspecting the open port distribution, we find a strong correlation for 10~MikroTik~ports~\cite{ceron2020mikrotik}.
OS and product information collected by Shodan confirm our observations becasuse the most common tags specify MikroTik.
Censys data confirms our results and also identifies the hosts as MikroTik devices.

\paragraph{Distribution in \texttt{/24} Prefixes}
We map each transparent forwarder to a (non-overlapping) covering \texttt{/24} IP~prefix and count the number of forwarders per prefix.
If all IP~addresses of a prefix reply to our transparent forwarder scans, we may assume that these replies are initiated by a single device (\eg some kind of middlebox that serves the whole prefix).
In contrast, for sparsely populated prefixes, we may assume multiple deployed devices (\eg several CPE devices that serve differnet customers).

\begin{figure}[b!]
  \begin{center}
  \includegraphics[width=1\columnwidth]{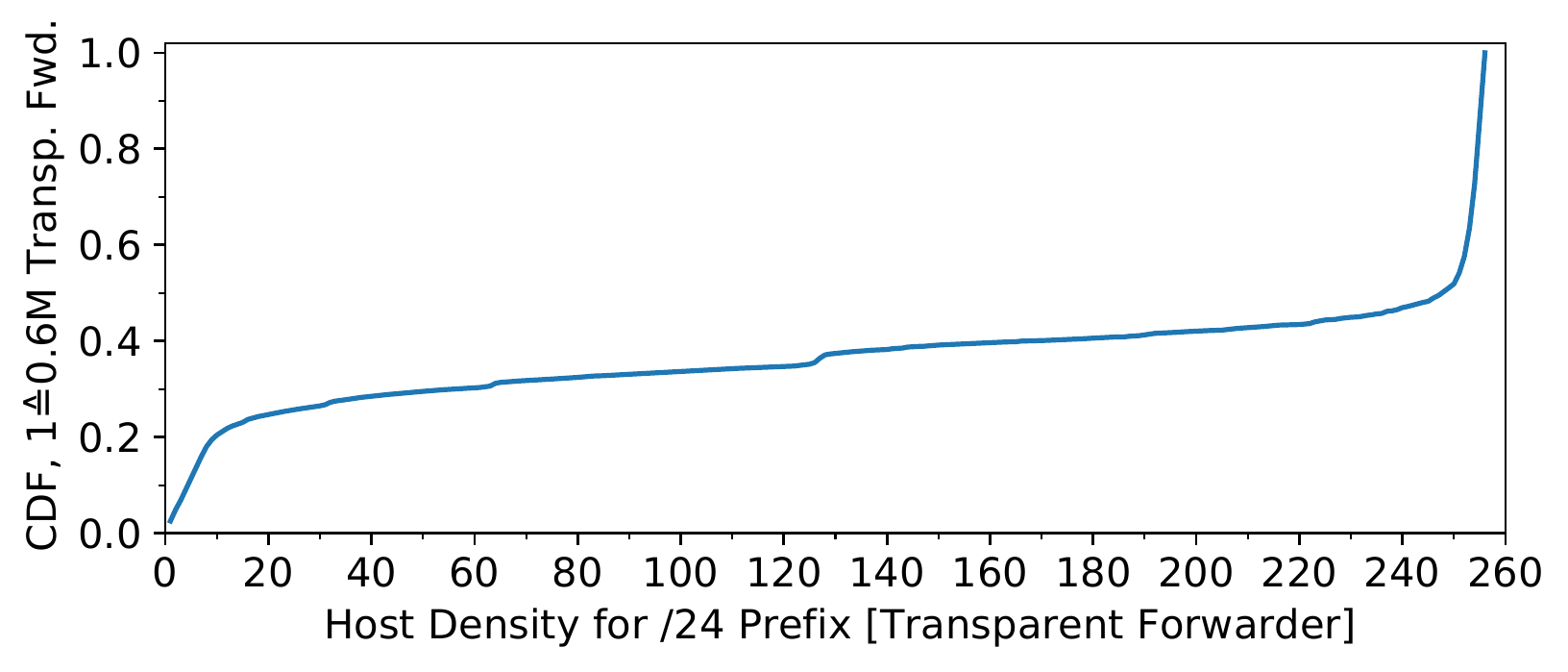}
      \caption{We map all transparent forwarders to a covering \texttt{/24} prefix. Some transparent forwarders belong to individual end customers, others may serve a larger network.}
  \label{fig:cdf_forwarder_prefixes}
  \end{center}
\end{figure}

41k distinct IP~prefixes cover 0.6M transparent forwarders.
\autoref{fig:cdf_forwarder_prefixes} shows the distribution of the number of transparent forwarders in each \texttt{/24}~prefix.
Overall, we observe a mixed picture.
26\% of all transparent forwarders are located in sparsely populated prefixes ($\leq25$~transparent forwarders in a \texttt{/24} prefix), and 36\% in completely populated prefixes ($\geq254$~~transparent forwarders in a \texttt{/24} prefix).
Only 806 prefixes are completely populated.
In those cases, we argue that a CPE device serves as a router for larger networks instead of a single end customer.
Using CPE devices outside of individual DSL/cable customers is not uncommon because CPE devices are cheap and some implement routing protococols (\eg MikroTik even BGP). 
In any case, whether the transparent forwarder function runs on a device that serves a single end customer or a larger network, our results hold, the transparent forwarder interacts as an ODNS component and uses the resolvers we observed.

\end{document}